\documentclass{iopart}
\input epsf

\def \D#1{{D$_#1$}}

\begin{document}

\title{Optical qubit generation by state truncation using an
experimentally feasible scheme}

\author{
{\rm \c{S}AHIN KAYA \"OZDEMIR$^{\dag}$\footnote[2]{E-mail:
ozdemir@koryuw01.soken.ac.jp}, ADAM MIRANOWICZ$^{\dag\S}$,\\
MASATO KOASHI$^\dag$, and NOBUYUKI IMOTO$^{\dag\Vert}$}}

\address{
$\dag$ CREST Research Team for Interacting Carrier Electronics,
School of Advanced Sciences, Graduate University for Advanced
Studies (SOKEN), Hayama, Kanagawa 240-0193, Japan}

\address{
$\S$ Nonlinear Optics Division, Institute of Physics, Adam
Mickiewicz University, 61-614 Pozna\'n, Poland}

\address{
$\Vert$ NTT Basic Research Laboratories, 3-1 Morinosato Wakamiya,
Atsugi, Kanagawa 243-0198, Japan}

\date{\today}
\pagestyle{plain} \pagenumbering{arabic}
%\maketitle

\vspace{5mm}
\begin{abstract}
Generation of arbitrary superposition of vacuum and one-photon
states using quantum scissors device (QSD) is studied. The device
allows the preparation of states by truncating an input coherent
light. Optimum values of the intensity of the coherent light for
the generation of any desired state using the experimentally
feasible QSD scheme are found. \vspace{5mm}
\end{abstract}

%%%%%%%%%%%%%%%%%%%%%%%%%%%%%%%%%%%%%%%%%%%%%%%%%%%%%%%%%%%%%%%%%%%%%%%%%%%%
\section{Introduction}
In recent years, many experiments falling into the quantum domain
of optical fields have been performed in optics laboratories. The
field has received much attention motivated not only by the
excitement of studying the fundamentals of quantum mechanics, but
also by the potential use of quantum optical light for the
manipulation and transmission of information. Consequently, the
quantum engineering of light, which consists of generation and
measurement of non-classical states of the optical field, has seen
a rapid development \cite{JMO}. Several schemes, which include the
Fock filtering scheme of D'Ariano {\em et al.} \cite{Dariano},
photon adding and displacing scheme of Dakna {\em et al.}
\cite{Dak99} and the optical state truncation of Pegg {\em et al.}
\cite{Peg98}--\cite{Kon00} have been proposed and studied.

The optical state truncation scheme, which is also referred to as
the quantum scissors device (QSD), was originally proposed for
preparing superposition of vacuum and one-photon state by
truncating a coherent light, and it has been modified by
Koniorczyk {\em et al.} \cite{Kon00} to generate superposition of
vacuum, one-photon and two-photon states. The original QSD scheme
is built by two beam splitters, one of which is fed by one-photon
in one input port whereas the second port is left at vacuum. One
of the output ports of this beam splitter is fed to the second
beam splitter where it is mixed with the coherent light. The
output modes of the second beam splitter are detected and the
condition in which one photon is detected in one of the modes and
none in the other mode corresponds to the conditional preparation
of vacuum and one-photon states at the output of the first beam
splitter. Paris \cite{Paris00} has modified this scheme by
replacing each beam splitter by a Mach-Zehnder interferometer and
proposed a fully interferometric correspondent of the QSD in which
relative weights of the vacuum and single-photon states can be
tuned by varying the internal phase-shift of the interferometers.

In a recent study \cite{Ozdemir01}, we have proposed an
experimental scheme for the practical realization of the QSD
scheme for state truncation taking into account the realistic
description of single-photon-state generation and photon counting
detectors. In this paper, that study is extended to include
generation of arbitrary superposition of vacuum and one-photon
states and to discuss the trade-off between the fidelity and
relative weights of the vacuum and one-photon states in the
superposition. The effect of detector efficiency is described and
the rate of preparation of the desired state is discussed. This
analysis goes beyond the analyses of \cite{Peg98,Bar99} by
considering the realistic descriptions of photodetection and
single-photon generation.

%%%%%%%%%%%%%%%%%%%%%%%%%%%%%%%%%%%%%%%%%%%%%%%%%%%%%%%%%%%%%%%%%%%%%%%%%%%%
\section{Experimental scheme for qubit generation}

The device proposed is schematically depicted in Fig.1. This
scheme is based on the ideas developed in
\cite{Rarity2000,Rarity1}. It consists of a parametric down
conversion crystal as the single-photon source, conventional
photon counters for conditional measurement and 50:50 beam
splitters for generation of entangled photon number states (BS1)
and for the mixing of coherent state with the entangled state
(BS2). The overall input to the QSD scheme is
\begin{equation}\label{N01}
\hat{\rho}_{{\rm in}1}=\hat{\rho}_{(a_1,c_1)} \otimes
|0\rangle_{a_{2} {a_{2}}} \langle0|\otimes |\alpha\rangle_{b_{3}
{b_{3}}} \langle \alpha|
\end{equation}
where $|\alpha\rangle_{b_{3}}$ is the coherent state to be
truncated to prepare the desired qubit state, and
$\hat{\rho}_{(a_1,c_1)}$ is a mixed state density operator
obtained at the output of the parametric down conversion after
averaging over all possible phases and is given as
\begin{equation}\label{N02}
\hat{\rho}_{(a_1,c_1)}=(1-\gamma^{2}) \Big[ |00 \rangle\langle 00
| + \gamma^{2}|11 \rangle\langle 11|+ \gamma^{4}|22 \rangle\langle
22|+ \cdots\Big]_{(a_1,c_1)}
\end{equation}
where $\gamma^{2}$, typically $\sim 10^{-4}$ \cite{Bouw97},
corresponds to the rate of one photon pair generation per pulse of
the pump field. The action of a beam splitter (BS1 and BS2) is
represented by a unitary operator $\hat{U}$. Given the input
density operator $\hat{\rho_1}$ of a lossless 50:50 beam splitter,
the output can be calculated using $\hat{U}^{\dagger}
\hat{\rho}_{1}\hat{U}$. $\hat{U}$ is conventionally given by
%%%%%%%%%%%%%%%%%%%%%%%%%%%%%%%%%%%%%%%%%%%%%%%%%%%%%%%%%%%%%%%%%%%%%%%%%%%%
%figure 1.
\vspace{5mm}
\begin{figure}[ht]
\vspace*{0mm}\hspace*{10mm} \epsfxsize=12cm
\epsfbox{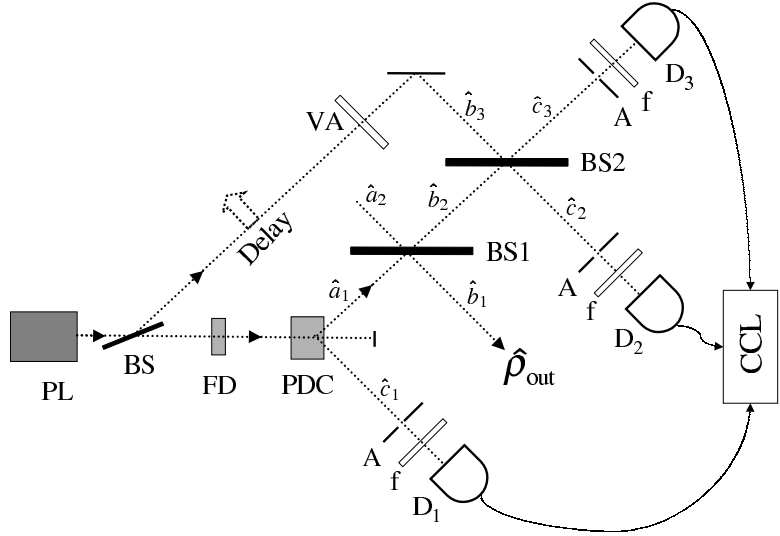} \vspace*{-15mm}%
\caption{Experimental QSD scheme for state truncation and qubit
state preparation: PL -- pulsed laser; FD -- frequency doubler;
PDC -- parametric down conversion crystal; VA -- variable
attenuator; A -- aperture; f -- narrow band filter; CCL --
coincidence counter and logic; BS, BS1, and BS2 are beam
splitters; and \D1, \D2, and \D3 are photon-counting
detectors.}\label{fig01}
\end{figure}
%----------------------------------------------------------------------
\begin{eqnarray}
\hat{U}&=&\exp[-i(\psi_{t}-\psi_{r})\hat{L}_{3}]
\exp[-i2\theta\hat{L}_{2}]\exp[-i(\psi_{t}+\psi_{r})\hat{L}_{3}],
 \label{N03}
\end{eqnarray}
where
%----------------------------------------------------------------------
\begin{eqnarray}
\hat{L}_{2}=\frac{1}{2i}(\hat{a}_{1}^{\dagger}\hat{a}_{2}
-\hat{a}_{2}^{\dagger}\hat{a}_{1}), \quad
\hat{L}_{3}=\frac{1}{2}(\hat{a}_{1}^{\dagger}\hat{a}_{1}
-\hat{a}_{2}^{\dagger}\hat{a}_{2}), \label{N04}
\end{eqnarray}
with $\theta=\pi/4$. $\psi_t=0$ and $\psi_r=\pi/2$ correspond to
the phases of the beam-splitter transmission and reflection
parameters, respectively. After the action of the beam splitters
(BS1 and BS2) on the input state $\hat{\rho}_{{\rm in}1}$, the
output state before the photodetection process becomes
\begin{eqnarray}
\hspace*{-5mm} \hat{\rho}_{(c_{1},c_{2},c_{3},b_{1})} =
\hat{U}_{2}^{\dagger}\hat{U}_{1}^{\dagger}{\big{(}}\hat{\rho}_{(a_1,c_1)}
\otimes|0 \rangle_{a_{2}}\,_{a_{2}}\!
\langle0|\otimes|\alpha\rangle_{b_{3}}\,_{b_{3}}\!\langle\alpha|{\big{)}}
\hat{U}_{1}\hat{U}_{2}.\label{N05}
\end{eqnarray}
The normalized truncated output state density operator at mode
${\hat b}_{1}$ of BS1 after the conditional measurement of
coincidence detection (``click") at detectors D1 and D2 and
anti-coincidence (``no-click") at D3 is obtained by
%----------------------------------------------------------------------
\begin{eqnarray}
\hat{\rho}_{\rm out}=\frac{{\rm{Tr}}_{(c_{1},c_{2},c_{3})}
(\Pi_{1}^{c_{1}} \Pi_{1}^{c_{2}} \Pi_{0}^{c_{3}}
\hat{\rho}_{(c_{1},c_{2},c_{3},b_{1})})}
{{\rm{Tr}}_{(b_{1},c_{1},c_{2},c_{3})}
(\Pi_{1}^{c_{1}}\Pi_{1}^{c_{2}} \Pi_{0}^{c_{3}}
\hat{\rho}_{(c_{1},c_{2},c_{3},b_{1})}
)} \, , \label{N06}
\end{eqnarray}
where $\Pi_{1}^{c_{1}}$, $\Pi_{1}^{c_{2}}$, and $\Pi_{0}^{c_{3}}$
are elements of the positive-operator-valued measures (POVMs) for
the detectors \D1, \D2, and \D3, respectively, with $0$ and $1$
corresponding to the number of clicks recorded at the detectors.
The measurement by conventional photon counters is described by
the following two-value POVM
\begin{eqnarray}
\Pi_{0} &=& \sum_{m=0}^\infty (1-\eta)^{m}|m\rangle \langle m|,
\nonumber\\
\Pi_{N\geq1} &=& 1-\Pi_{0}\, .\label{N07}
\end{eqnarray}
with $\eta$ being the quantum efficiency of the detectors. The
effect of mean dark count $\nu$ of the detector in the POVM has
been neglected because in a previous study, it was shown that when
$\nu\ll\gamma^{2}$, which can be achieved in experiments using
commercially available detectors, dark count rate does not have a
significant effect on the fidelity and efficiency of the
truncation process.

Due to non-ideal detection and non-ideal single-photon generation
(output of parametric down conversion process contains vacuum and
higher number of photons together with a low rate of single-photon
pair), the conditional output state $\hat{\rho}_{\rm out}$ is not
a pure state. However, as was shown in \cite{Ozdemir01}, there are
regimes of intensity of the input coherent state and the detector
parameters where the output state approaches the superposition
state of vacuum and one-photon with non-vanishing probability. To
compare $\hat{\rho}_{\rm out}$ with the desired qubit state of the
form $|\psi_{\rm desired}
\rangle=N[c_{0}|0\rangle+c_{1}|1\rangle]$ with $N$ being the
normalization constant, we consider the fidelity
$F=\langle\psi_{\rm desired}|\hat{\rho}_{\rm out}|\psi_{\rm
desired}\rangle$.

%%%%%%%%%%%%%%%%%%%%%%%%%%%%%%%%%%%%%%%%%%%%%%%%%%%%%%%%%%%%%%%%%%%%%%%%%%%%
\section{Preparation of qubit states with the experimental QSD scheme}

In the QSD scheme there are two kinds of free parameters, namely
the intensity $|\alpha|^{2}$ of the coherent light and the beam
splitter parameter (reflectance and transmittance), that can be
tuned to properly set the relative weights of the vacuum and
one-photon states in the superposition of $N[c_{0}|0\rangle
+c_{1}|1\rangle]$. The tuning of the beam splitter parameters can
be realized by using the interferometric scheme of Paris
\cite{Paris00}. This scheme makes the setup more complicated and
introduces the problem of controlling the stability and balance of
the interferometers. The adjustment of the $|\alpha|^{2}$ can be
realized without introducing additional complexity to the setup
except a controllable tuning of the intensity of the coherent
light. However, this may also constitute a challenge for the
preparation of some specific qubit states which needs very fine
tuning of the intensity.

In this study, our aim is to optimize the intensity $|\alpha|^{2}$
of the input coherent light to generate an arbitrary qubit state
with the highest fidelity with non-vanishing probability of the
state generation. In that case, $BS1$ and $BS2$ are chosen as
$50:50$ beam splitters because they give the highest probability
and fidelity for state truncation using the ideal QSD scheme
\cite{Peg98,Ozdemir01}. The highest fidelity in generating
arbitrary qubit states from a coherent state can be achieved if
arg($\alpha$)=arg($c_{1}$), which is assumed to be the case in
this study.

In the following, the results of numerical simulations for the
proposed experimental QSD scheme are presented. The simulations
were performed for different $|\alpha|^{2}$ and $\eta$ using the
POVM given by (\ref{N07}). Parametric down conversion output was
used in the simulations in the form of (\ref{N02}) and
$\gamma^{2}$ parameter was taken as $4\times10^{-4}$ for which the
ratio of one-photon pair generation to that of the two-photon pair
generation is $O(10^{4})$ resulting in a low probability of having
contributions from higher number of photons at the output mixed
state.

%%%%%%%%%%%%%%%%%%%%%%%%%%%%%%%%%%%%%%%%%%%%%%%%%%%%%%%%%%%%%%%%%%%%%%%%%%%%
\subsection{Equally weighted superposition of vacuum and one-photon states}

\begin{figure}[ht]
\vspace*{-7mm}\hspace*{0mm} \epsfxsize=6.5cm
\epsfbox{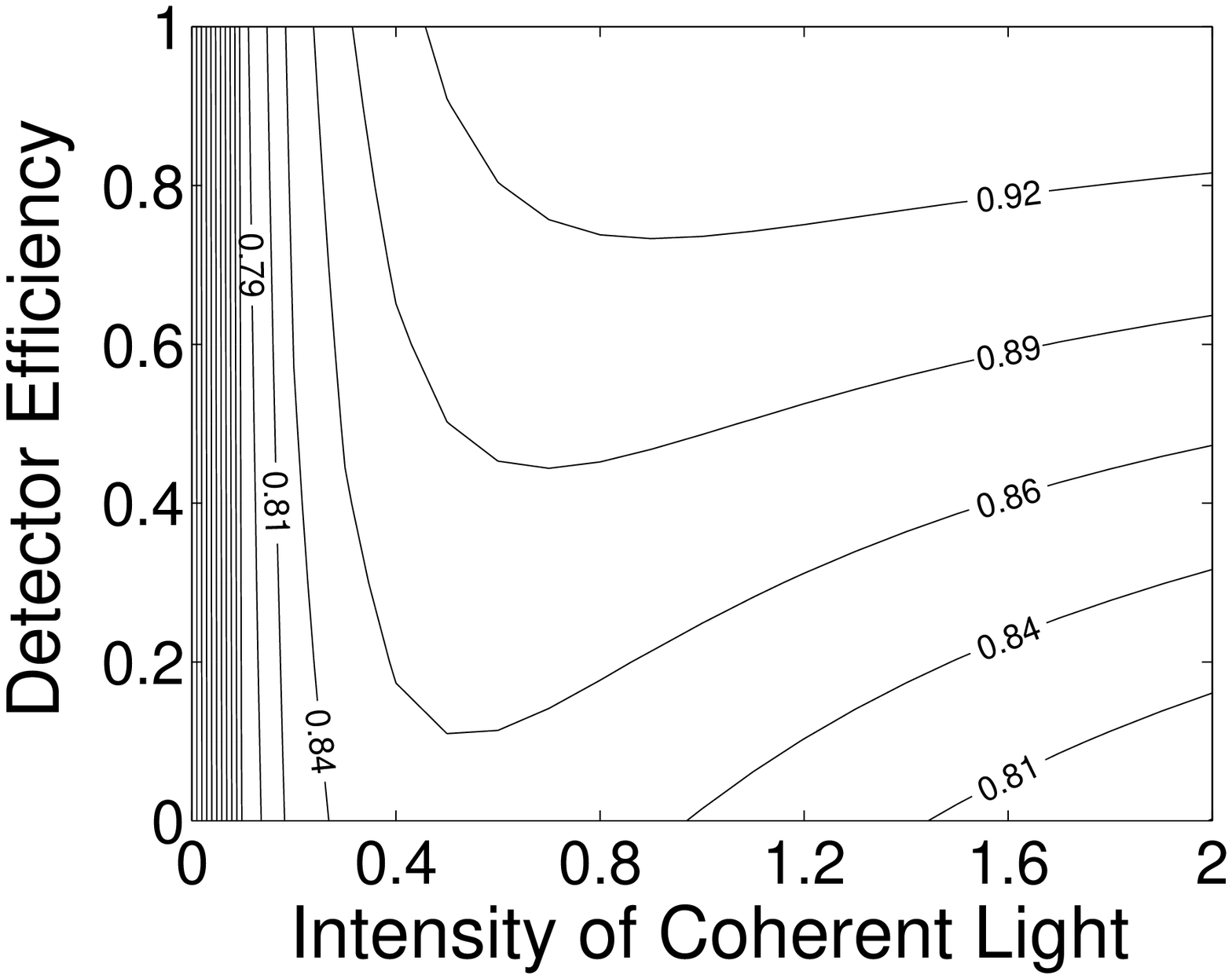}\hspace*{0mm} \epsfxsize=6.5cm
\epsfbox{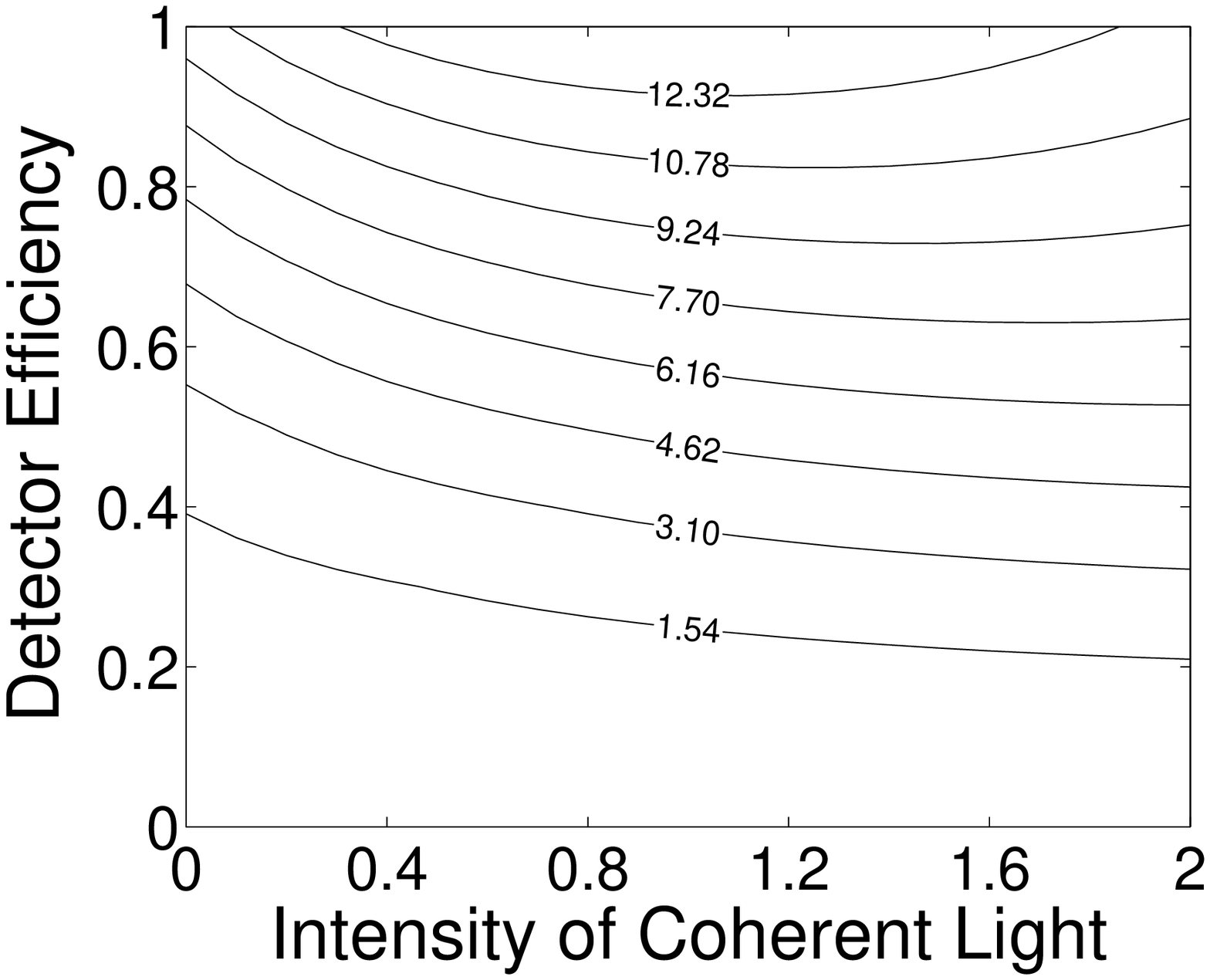} \vspace{-3mm}%
\caption{Effect of intensity of the input coherent light
$|\alpha|^{2}$ and detector efficiency $\eta$ on the fidelity $F$
(left) and the rate $R$ of correct detection event (right) of the
balanced superposition state generation. Constant $F$ and $R$
curves are depicted in the figure. $R$ curves are scaled with
$100$ for clarity.}\label{fig02}
\end{figure}

Effects of detection efficiency and the intensity of the coherent
light on the preparation of the qubit state which consists of
equally weighted vacuum and one-photon components of the form
${\big{[}}~|0\rangle+|1\rangle~{\big{]}}/\sqrt{2}$ are analyzed .
The results of these simulations are depicted in figure
\ref{fig02}. The relation between $\eta$ and $|\alpha|^{2}$ to
achieve a given fidelity is clearly seen in this figure. For
$|\alpha|^{2} \leq 0.36$, fidelity is less than $0.9$ for any
value of detector efficiency. At much smaller values of intensity,
fidelity is not affected by the detector efficiency. At a constant
$|\alpha|^{2}$, the number of correct detection events increases
with increasing detector efficiency. It is seen that equally
weighted superposition of vacuum and single-photon states can be
prepared with high fidelity even for $\eta$ as low as $0.1$
provided that the intensity of the coherent light is chosen
properly. For $\eta=0.5$, the optimum value for $|\alpha|^{2}$ is
$0.72$ for which the fidelity of the prepared state to the desired
one becomes $0.89$. Such a detection and state preparation can be
achieved with a rate of 4533 s$^{-1}$. For each detection
efficiency, the optimized value of \linebreak

%%%%%%%%%%%%%%%%%%%%%%%%%%%%%%%%%%%%%%%%%%%%%%%%%%%%%%%%%%%%%%%%%%%%%%%%%%%%
%figure 2.
\vspace{5mm}
\vspace*{-2mm} \noindent $|\alpha|^{2}$ and the maximum attainable
fidelity are different. Increasing $\eta$ from $0.5$ to $0.7$ will
increase the value of $|\alpha|^{2}$ to $1.06$ and will change the
fidelity only by $0.11\%$. On the other hand, the generation rate
of such a state is almost doubled to a value of 8524 s$^{-1}$.

%%%%%%%%%%%%%%%%%%%%%%%%%%%%%%%%%%%%%%%%%%%%%%%%%%%%%%%%%%%%%%%%%%%%%%%%%%%%
\subsection{Arbitrary superposition of vacuum and one-photon states}

\begin{figure}[ht]
\vspace*{0mm}\hspace*{0mm} \epsfxsize=13cm
\epsfbox{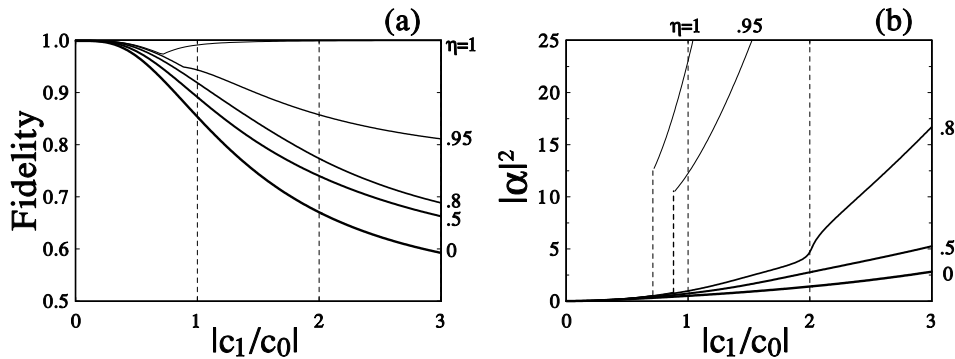} \vspace*{-4mm}%
\caption{Optimized $|\alpha|^{2}$ for preparing qubit states of
arbitrary $|c_{1}/c_{0}|$ and the corresponding maximum fidelity
for various detector efficiencies.}\label{fig03}
\end{figure}

The intensity $|\alpha|^{2}$ of the coherent light has been
optimized to give the maximum fidelity to prepare an arbitrary
superposition of vacuum and one-photon state in the form of
\begin{equation}
|\psi_{\rm desired}\rangle=\frac{c_{0}|0\rangle
+c_{1}|1\rangle}{\sqrt{|c_{0}|^{2}+|c_{1}|^{2}}}. \label{N08}
\end{equation}
\noindent Figure \ref{fig03} depicts the values of the optimized
$|\alpha|^{2}$ and the maximum value of fidelity that can be
achieved for any desired state of $|c_{1}/c_{0}|$ with that value
of $|\alpha|^{2}$ for different detector efficiencies. It has been
understood that an optimum $|\alpha|^{2}$, which will yield a
maximum fidelity with non-zero preparation rate, can always be
found for any arbitrary desired state. This optimum value of
$|\alpha|^{2}$ depends on the detector efficiency $\eta$ and
$|c_{1}/c_{0}|$. When $|c_{1}/c_{0}|\leq0.4$, a fidelity value of
$\sim 0.99$ can be achieved with the optimized $|\alpha|^{2}$ for
$\eta\geq0.5$. The fidelity of the qubit state preparation depends
significantly on the relative weights of the vacuum and
single-photon states in the superposition. If the vacuum component
is dominant in the desired state, then high fidelity values
greater than $0.9$ can be achieved even for smaller detector
efficiencies. But when the one-photon component becomes dominant,
the fidelity of the state preparation starts decreasing. For
$|c_{1}/c_{0}| =0.5$, $F\simeq0.98$ at the optimized value
$|\alpha|^{2} \simeq0.21$ and $\eta=0.5$, however when the desired
state is $|c_{1}/c_{0}|=2$, the fidelity drops to 0.74 at the
optimized value of $|\alpha|^{2}\simeq2.7$. If $\eta$ is
increased, fidelity increases to only $\sim0.77$ at $\eta=0.7$,
and to $\sim0.8$ at $\eta=0.9$. All of these arbitrary qubit
states can be generated with non-zero preparation rate. This
preparation rate increases with increase in $|c_{1}/c_{0}|$ up to
some maximum value which depends on $\eta$ and then starts
decreasing when one-photon state becomes more dominant in the
superposition.

It is  observed that when detector efficiency is in the range of
$1.0\leq\eta\leq0.9$, for some values of $|c_{1}/c_{0}|$, two
different values of optimum $|\alpha|^{2}$ at which fidelity takes
the \linebreak
%%%%%%%%%%%%%%%%%%%%%%%%%%%%%%%%%%%%%%%%%%%%%%%%%%%%%%%%%%%%%%%%%%%%%%%%%%%%
%figure 3.
\vspace{5mm}

\noindent same maximum value can be found (see figure
\ref{fig03}(b)). Although fidelity reaches the maximum value for
these two values of $|\alpha|^{2}$, the corresponding rates of the
preparation of the state are different. As it can be seen in
figure \ref{fig03}, for the preparation of the qubit state with
$|c_{1}/c_{0}|\simeq1.145$ using detectors of $\eta=0.9$, the
maximum fidelity of $\sim0.906$ can be obtained at
$|\alpha|^{2}\simeq1.464$ and $|\alpha|^{2}\simeq8.644$ for which
the state preparation rates are 11869s$^{-1}$ and 561s$^{-1}$,
respectively. In the same way, for $|c_{1}/c_{0}|\simeq0.712$, if
we increase $\eta$ to unity, the state preparation rate will
become 13330s$^{-1}$ and 35s$^{-1}$ at $|\alpha|^{2}\simeq0.548$
and $|\alpha|^{2}\simeq12.532$, respectively, with the same
maximum attainable fidelity of $\sim0.973$.

It must be noted that optimizing the intensity of the coherent
light to obtain the maximum fidelity for a desired qubit state
does not necessarily mean that the system is maximized for the
highest preparation rate, too. The cost of maximizing the fidelity
is the decrease in the preparation rate and vice versa. At
$\eta=0.7$, to prepare a qubit state having $|c_{1}/c_{0}|=0.4$
with the highest attainable fidelity of $0.992$, the optimum
$|\alpha|^{2}=0.152$ must be used which will result in preparation
rate of 5790s$^{-1}$. However, if the system is optimized to have
the highest preparation rate, an optimum value of
$|\alpha|^{2}=1.538$ needs to be used. This will result in  a
fidelity of $0.859$ which is much lower than that of the former
case. At $\eta\geq0.7$, only for the case of balanced
superposition state, the values of fidelity and state preparation
rate obtained when the system is optimized for the highest
fidelity are very close to those obtained when optimization is
made for the highest preparation rate.

\begin{figure}[ht]
\vspace*{0mm}\hspace*{0mm} \epsfxsize=13cm
\epsfbox{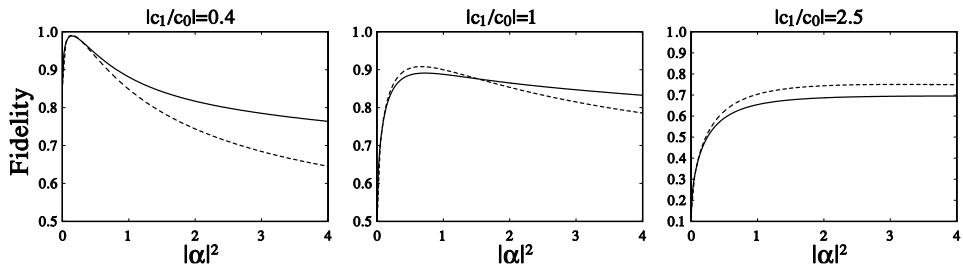} \vspace*{-5mm}%
\caption{Comparison of the dependence of fidelity for the proposed
experimental scheme and the ideal QSD scheme \cite{Bar99} when
they are used for preparing arbitrary qubit states. Solid line and
dotted line correspond to experimental and the ideal QSD schemes,
respectively.}\label{fig04}
\end{figure}

Figure \ref{fig04} depicts the comparison of the ideal QSD scheme
of Barnett {\em et al.} \cite{Bar99} and the proposed experimental
scheme for $|\alpha|^2\leq4$ and $\eta=0.5$. We have not
considered the larger values of $|\alpha|^2$, because at these
higher values of light intensity, even though in some cases,
fidelity has a higher value, the state preparation rate approaches
very small values which are not experimentally feasible. The
comparison has been done in three different regions of
$|c_{1}/c_{0}|$ for various $\eta$: (a) For $|c_{1}/c_{0}|<1.0$,
which means vacuum dominated qubit state, fidelity increases with
increasing $|\alpha|^2$ until it reaches the maximum value for
both schemes. After that maximum value, fidelity starts decreasing
with respect to $|\alpha|^2$, however the experimental scheme
overwhelms the results obtained in \cite{Bar99} for the ideal
scheme and has much higher fidelity values. It is seen in figure
\ref{fig04} that for $|c_{1}/c_{0}|=0.4$, the fidelity values for
both schemes are very close to each other until
$|\alpha|^2\simeq0.28$ beyond which fidelity of the ideal scheme
decreases \linebreak

%%%%%%%%%%%%%%%%%%%%%%%%%%%%%%%%%%%%%%%%%%%%%%%%%%%%%%%%%%%%%%%%%%%%%%%%%%%%
%figure 4.
\vspace{5mm}
 \noindent
more rapidly with increasing values of $|\alpha|^2$. (b) For the
case of balanced superposition state having $|c_{1}/c_{0}|=1.0$,
ideal QSD scheme has higher value of fidelity than the
experimental scheme up to a critical value of
$|\alpha|^2\simeq1.52$ after which the experimental scheme starts
to give slightly higher fidelity values; this difference is not as
pronounced as the difference seen in case (a). (c) For
$|c_{1}/c_{0}|>1.0$ corresponding to the one-photon-dominant qubit
states, the ideal QSD scheme prepares states with higher fidelity
than that of the proposed experimental scheme. When the total
number of photons incident on the detectors is large, the output
state approaches the vacuum state enhancing the advantage of the
experimental scheme.

%%%%%%%%%%%%%%%%%%%%%%%%%%%%%%%%%%%%%%%%%%%%%%%%%%%%%%%%%%%%%%%%%%%%%%%
\section{CONCLUSIONS}

The experimentally feasible optical state truncation scheme
proposed in \cite{Ozdemir01} has been analyzed for the preparation
of arbitrary qubit states of the form given by (\ref{N08}). State
preparation is based on truncating a coherent light to contain
only its vacuum and one-photon components. The relative weights of
the vacuum and one-photon states in the qubit state can be
adjusted by manipulating the intensity of the coherent light. It
is understood that the qubit states, which are dominated by the
vacuum component, can be prepared with high fidelity and
efficiency with the proposed scheme considering the realistic
descriptions of photodetection and single-photon generation.
Moreover, one can always find an optimum value for the intensity
of the coherent light which will maximize the fidelity. However,
the price for the maximization of the fidelity is the decrease in
the preparation rate of the state. This study reveals that the
original QSD scheme and the proposed experimental scheme can be
used not only for truncating an input coherent state but also for
generating arbitrary qubit states with proper manipulation of the
light intensity.

%%%%%%%%%%%%%%%%%%%%%%%%%%%%%%%%%%%%%%%%%%%%%%%%%%%%%%%%%%%%%%%%%%%%%%%%%%%%
\section*{Acknowledgments}

We thank Prof. Stephen M. Barnett for pointing us this problem and
for his invitation and hospitality during Quantum Electronics and
Photonics 15 (QEP-15) conference. We also thank Takashi Yamamoto
and Yu-xi Liu for their stimulating discussions. This work was
supported by a Grant-in-Aid for Encouragement of Young Scientists
(Grant No.~12740243) and a Grant-in-Aid for Scientific Research
(B) (Grant No.~12440111) by Japan Society for the Promotion of
Science.

%%%%%%%%%%%%%%%%%%%%%%%%%%%%%%%%%%%%%%%%%%%%%%%%%%%%%%%%%%%%%%%%%%%%%%%%%%%%

\vspace{5mm}
%%%%%%%%%%%%%%%%%%%%%%%%%%%%%%%%%%%%%%%%%%%%%%%%%%%%%%%%%%%%%%%%%%%%%%%%%%%%
{\setlength{\fboxsep}{10pt}
\begin{center}
\framebox{\parbox{0.75\columnwidth}{%
\begin{center}
to appear in
% a special QEP-15 issue of
the {\em Journal of Modern Optics}
% (Special Issue Editor: Prof. S. M. Barnett)\\
(spring 2002).
\end{center}}}
\end{center}

\end{document}